# Scaling up the software development process, a case study highlighting the complexities of large team software development.

*Mark Basham, Diamond Light Source, Harwell Science and Innovation Campus, Didcot OX11 0DE , mark.basham@diamond.ac.uk.*

## Abstract

Diamond Light Source is the UK's National Synchrotron Facility and as such provides access to world class experimental services for UK and international researchers.  As a user facility, that is one that focuses on providing a good user experience to our varied visitors, Diamond invests heavily in software infrastructure and staff.  Over 100 members of the 600 strong workforce consider software development as a significant tool to help them achieve their primary role.  These staff work on a diverse number of different software packages, providing support for installation and configuration, maintenance and bug fixing, as well as additional research and development of software when required.

This talk focuses on one of the software projects undertaken to unify and improve the user experience of several experiments.  The "mapping project" is a large 2 year, multi group project targeting the collection and processing experiments which involve scanning an X-ray beam over a sample and building up an image of that sample, similar to the way that google maps bring together small pieces of information to produce a full map of the world.  The project itself is divided into several work packages, ranging from teams of one to 5 or 6 in size, with varying levels of time commitment to the project. This paper aims to explore one of these work packages as a case study, highlighting the experiences of the project team, the methodologies employed, their outcomes, and the lessons learnt from the experience.

## Introduction

The Diamond mapping project [1] is a cross group and division project which has been running at Diamond Light Source [2] for the last 18 months.  Diamond is currently the home of 33 separate beamlines, each of which has its own scientific aim and dedicate scientific staff.  Most of these beamlines make use of the same software stack [3] but due hardware complexities and the challenging timescales of cutting edge research, many solutions are more custom than generic in nature.  Although many of the beamlines at Diamond have significantly different scientific aims, there are many similarities in the motion control, data acquisition and data analysis which can be exploited to create one suite of software tools which is universally applicable.  The mapping project's aim in particular is to unify the mapping style scans of five similar beamlines, with both hardware and software stacks.  This paper looks to show some of the challenges which occur with initial scale-up of a software support group, and then to show one specific area of the Mapping project, and how we tried to accelerate its progress.

## Support Groups and how they Scale-up

### Scale up at Diamond Light Source

In 2006 just before Diamond came online for first users, the ratio of Software developers in the Data Acquisition, Data Analysis and Beamline Controls groups to beamlines, was around 2 to 1, there were 2 dedicated members of software development staff for every beamline. One additional advantage of this dedication was that the software staff could get to fully understand the requirements, and intricate details of the work required as they were really part of the team. In 2016 however, with greater specialisation as is required during scaleup, the staff are spread between 3 groups rather than 2. Each group can no longer provide 1 member of staff per beamline, and this is exacerbated by the general requirements for bigger teams to have more core staff who are not supporting beamlines directly, but maintaining central infrastructure (Figure 1). Ultimately the result of this is that now one developer may well have to support multiple beamlines, sharing their time between them. This also has the effect that these developers are not able to integrate fully with the beamline team, which introduces a slight communications barrier making requirements gathering more complex.

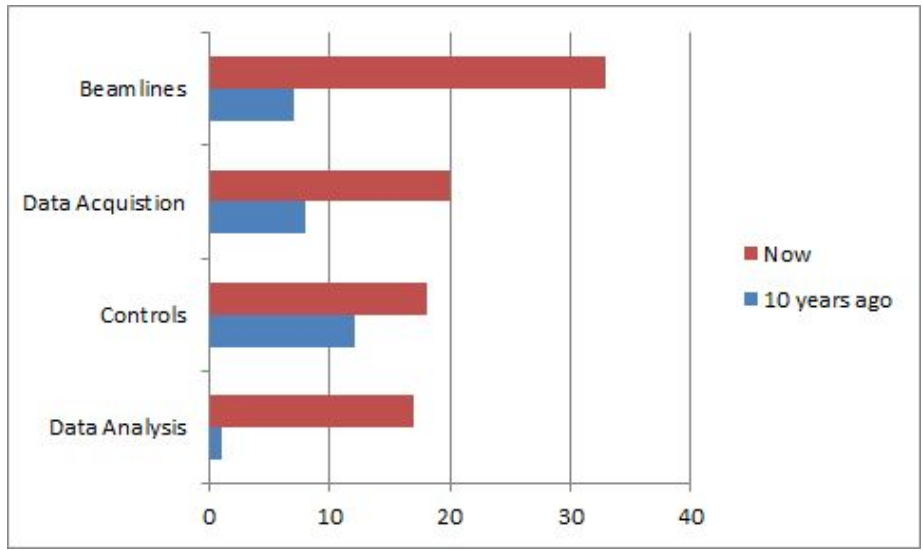

Figure 1 - The change in Beamline to Staff over the period 2006 - 2016

Central Software Providers

Given diamond has had centralised teams for all its core software, at first glance the mapping project may seem strange. With these central teams working together, on effectively the same software stacks, how do we end up with different solutions on each beamline. In the author's opinion this is due to the effect noted in Fred Brooks "The Mythical Man Month" [4] where he talks about talking a program and adding value.

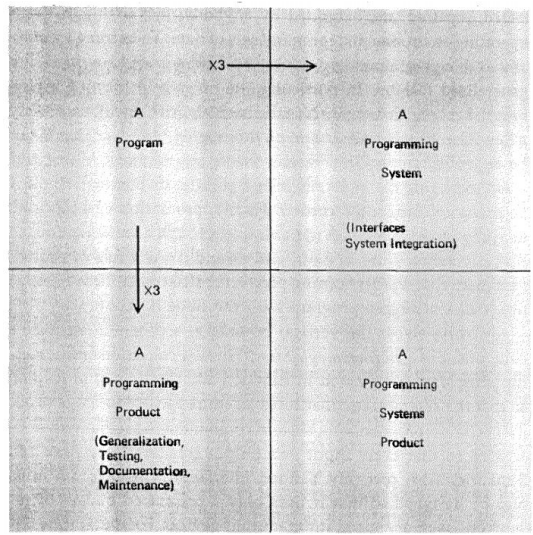

Figure 2 - Costs of generalizing programming. Figure from Figure 1.1 of Fred Brooks. The Mythical Man-Month. Pearson Education India; 1995 Sep 1

The added values which he talks about in Figure 2 are all good software engineering practices, such as testing, documentation, clear interfaces, using existing frameworks, etc.  What is noted however is that to achieve this level of complete integration and sustainability requires two sets of three-fold increase in time, hence a nine-fold increase in the effort required to achieve the goal.  At many places and especially at diamond, significant time restraints are on the software developers, and often as the last part in a project which can take multiple years, the final time time to deploy is shorter than would be liked to meet the challenging scientific requirements. With these time constraints in place it is easy to see how how the almost order of magnitude increase in time to truly make the software generic is a luxury which often cannot be afforded leading to an observed propagation of not fully integrated components for doing specific jobs. This in turn means that the next developer who needs to use one of these features, will often end up writing it from scratch rather than trying to reuse the existing code, as the time constraints prohibit significant refactoring.  It is only when a separate project is started with the specific aim of making a sustainable and fully integrated solution, such as the mapping project, that these elements become important enough to justify the additional cost in time and effort from the developers.

## Context Switching and Interruptions

Given that developers are often involved with more than one project, not only is their time shared between these beamlines, but there is added context switching time added to this.  Flow time as described in "Peopleware" [5] is essentially uninterrupted time, usually measured in full hours.  The premise is that for knowledge work, i.e work that requires the worker to use their experience and skills to achieve the goal, flow time is required for them to achieve progress. Context switching is one way that flow can be interrupted, but with careful planning it is easy enough for a developer to partition their time so that the work on one topic for half or a full day at a time.  The real issue is introduced with external (or unplanned) interruptions, where a developer is required to also interface with the users or other developers for support. Interruptions become even worse if the developer is dealing with multiple projects, as an interruption in the project you are working on currently is probably relevant, whereas an interruption from another project is almost certainly going to require you to context shift twice, once to the interrupting project, and then back to what you were doing.  If we simply assume that interruptions are the only time we will lose flow, the impact of that given a number of projects and the chance of interruptions is shown in Figure 3.

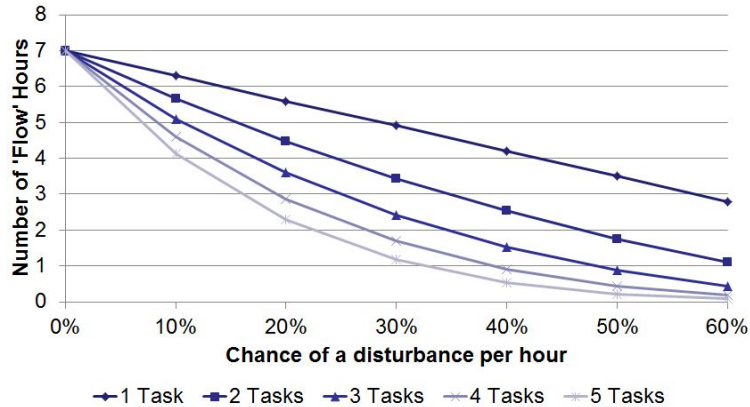

Figure 3 - Average number of "Flow Hours" per day given a number of tasks and a chance of interruption.

It is clear to see that as the number of tasks increases, there is a marked effect on the number of flow hours on average that a developer will have, even 3 tasks removes an additional hour of productive time per day with a 10% chance of interruption.

Requirements

As mentioned in the introduction, a key factor which occurred when the ratio of staff to beamlines decreased was the inability for the developers to be fully integrated with the beamline teams.  This leads to requirements being much more difficult to gather, as they are being gathered by an outsider rather than part of the team.  With new projects identifying the goals and what the end users really want from a project is complex and the Mapping Project suffered with this heavily at the start of the project.

Environment

An underestimated element for developer productivity is work environment, as detailed in the work by the Atlantic Systems Guild in their study of 166 software developers in "coding war games"[6].  The key finding of the paper is that although there was a difference in developer speed of the task of 10 times, i.e. the fastest developers finished the task in one hour, the slowest in ten, this was not related to pay or experience, rather the working environment the developer was in.  The study found that people sharing the same working environment had much more closely matching scores and on further investigation office space seemed to be the key criteria.  Crudely put developers from the top quartile had an average of 78 sq ft. of office space, as opposed to those in the bottom quartile who had an average of 46 sq ft.  Open plan office space is common in large organisations as an effective way of housing staff, but the average perceived area of space for members of the mapping project is 43 sq ft.

## Specific Case Study

The specific area which we will focus this case study on is work package 4 of the mapping project, the "Malcolm" middle layer between the main controls software and the main data acquisition software. Although the purpose of this work package is not that important, about 1 year into the project it was found to be about 5 months behind schedule, and looking to miss the main deployment dates which were required. As this was a critical part of the project, Diamond decided to try a more radical solution to the problem.

First we identified a team of people who could achieve the goals of the project, and then worked with their line managers to remove all other responsibilities from them for a month. This was not practical, so a compromise was made that they still spent one hour after lunch as their main contact time. However even with this compromise it meant that 6 hours of their day were on one project, and effectively interruption free.

Second we set up a room in a separate part of the building for them to use (Figure 4). The benefits of this were significant and far more than were initially expected. Firstly it made the whole thing seem more real, and the people set up in the room felt the "Black Team"[7] effect of being picked for this specific big task. Secondly they were able to make it their own, which meant that most of the walls were covered in sticky notes for planning, and the whiteboards were always full of architecture diagrams. Thirdly it increased their desk space by about 70% over that of the open plan(Moving them close to top quartile developers[6]), and also reduced off topic noise .

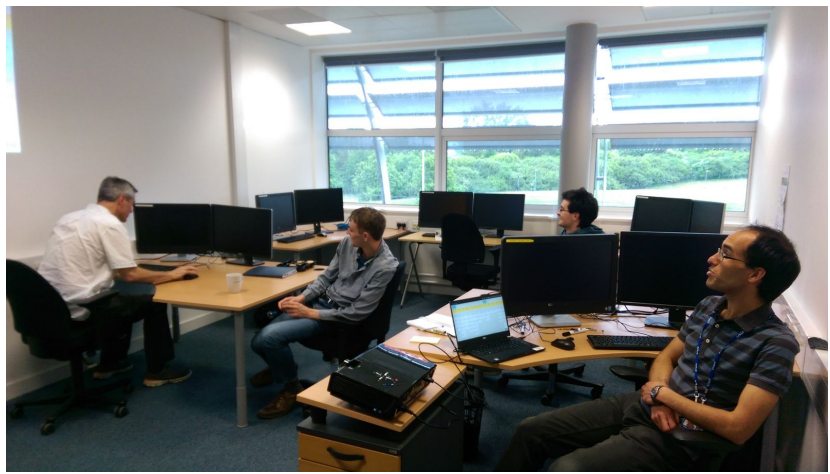

Figure 4

Finally we Adopted the "User Story Mapping" [8] methodology for making sure our requirements were right and our design was sound. This method gets you to tell the user story of how the product needs to work from the perspective of all the different users who would need to use it. You then flesh out the individual items of the story elements with smaller stories until you feel that you have understood the problem fully. Once you have this map, the team can then prioritise elements to achieve, and then finally pick these items to break down into developer specific tasks. Figure 5 shows the midway point of the malcolm team's story map.

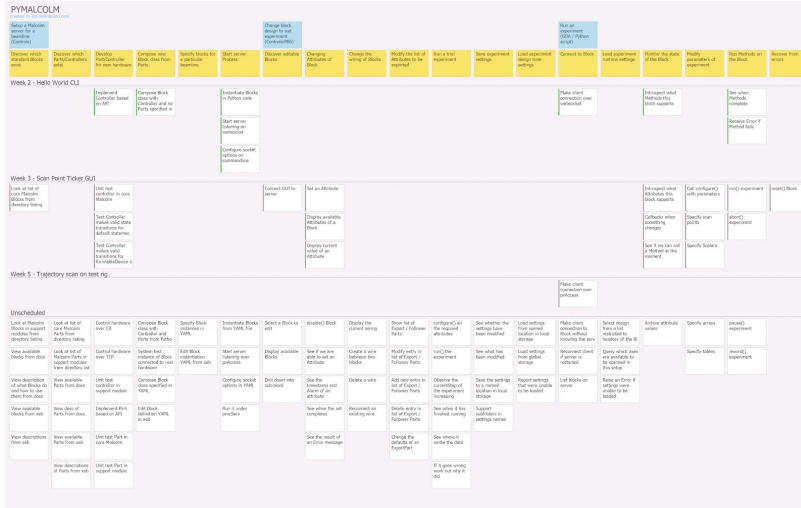

Figure 5 - the story map halfway through the trial, showing some completed work, and much left to be done in the backlog.

## Conclusion

After the 2 month period of the trial was over we reevaluated the work package to see what the current status was. The period of the trial had successfully caught the project up to only 1 month behind schedule, from 5 months before the trial. It had done so whilst producing a very high level of testing and documentation as can be seen from its github pages [9] reducing the risks associated with the work package. The trial has been considered a success by all participants, and although there are some difficulties associated with setting up such an environment, the productivity achieved was well worth the effort.

## Acknowledgements

The team at Diamond who helped with this work, Alun Ashton, Tom Cobb, Giles Knapp, Charles Mita, Gary Yendall.